\documentclass[aps,prd,floats,nofootinbib,preprintnumbers]{revtex4}
\usepackage{epsfig}
\usepackage{amsmath}
\usepackage{amssymb}
\usepackage{amsthm}

\begin{document}

\preprint{NUHEP-TH/06-06}

\title{Neutrino Phenomenology of Very Low-Energy Seesaws}

\author{Andr\'e de Gouv\^ea}
\affiliation{Northwestern University, Department of Physics \& Astronomy, 2145 Sheridan Road, Evanston, IL~60208, USA}

\author{James Jenkins}
\affiliation{Northwestern University, Department of Physics \& Astronomy, 2145 Sheridan Road, Evanston, IL~60208, USA}

\author{Nirmala Vasudevan}
\affiliation{Northwestern University, Department of Physics \&
Astronomy, 2145 Sheridan Road, Evanston, IL~60208, USA}

\begin{abstract}
The Standard Model augmented by the presence of gauge-singlet right-handed neutrinos
proves to be an ideal scenario for accommodating nonzero neutrino masses. Among the new parameters of this ``New Standard Model'' are right-handed neutrino Majorana masses $M$. Theoretical prejudice points to $M$ much larger than the electroweak symmetry breaking scale, but it has recently been emphasized that all $M$ values are technically natural and should be explored. Indeed, $M$ around $1-10$~eV can accommodate an elegant oscillation solution to the LSND anomaly, while other $M$ values lead to several observable consequences.
We consider the phenomenology of low
energy ($M\lesssim 1$~keV) seesaw scenarios.  By
exploring such a framework with three right-handed neutrinos, we can consistently fit all
oscillation data -- including those from LSND -- while partially addressing several astrophysical puzzles,
including anomalous pulsar kicks, heavy element nucleosynthesis in supernovae, and the existence of
warm dark matter.  Furthermore, low-energy seesaws -- regardless of their relation to the LSND anomaly -- can also be tested by future tritium beta-decay experiments, neutrinoless double-beta decay searches, and other observables. We estimate the sensitivity of such probes to $M$.
\end{abstract}

\maketitle

\setcounter{equation}{0} \setcounter{footnote}{0}
\section{Introduction}
\label{sec:intro}

Neutrinos have provided our first glimpse into physics beyond the
Standard Model of elementary particles (SM).  Neutrino oscillation
experiments have given us unambiguous evidence that the three active
neutrinos have mass and mix  (see \cite{neutrino_review} for recent
reviews of neutrino theory and phenomenology).  As most important
discoveries, the results of these experiments have raised more
questions than they answered. Even from our limited knowledge of the
neutrinos, it is clear that their properties, including  sub-eV
masses and large mixing, are quite different from their charged
fermion counterparts.  The true explanation of this puzzling
behavior likely rests on the fact that neutrinos are the only known
fundamental neutral fermions, but the exact reason behind this is
still open to speculation. Neutrino masses, as deduced from
oscillation experiments, hint at the existence of right-handed
neutrino states, lepton number violation, new sources of CP
violation, as well as a new energy scale.

The seesaw mechanism \cite{seesaw} is an appealing way to generate
the observed neutrino masses and lepton mixing matrix. The idea is simple.
Add an arbitrary number of singlet fermion states to the SM matter content.
The triviality of their quantum numbers allows them to have Majorana masses of magnitude
$M$, as well as couple to the $SU(2)_L$ lepton and Higgs doublets.
The latter vertices become Dirac mass
terms of magnitude $\mu$ after electroweak symmetry breaking. The standard theoretical
prejudice is
that the Dirac masses are of order the charged fermion masses, while
the Majorana masses are at some very high energy scale, $M\gg 100$~GeV.
If this is indeed the case, the resulting propagating neutrino degrees of freedom separate into two
quasi-decoupled groups: mostly active states with very small masses $m\sim \mu^2/M$
suppressed by the new physics scale, and mostly sterile states with very large masses $M$. In this
scenario, the mostly right-handed states are not directly observable. Indeed, it is possible that if such a high-energy seesaw is realized in Nature, its only observable consequence is that the mostly active neutrinos have mass and mix.

 Of course, there is no direct evidence that $M$ -- which we refer to as the seesaw scale --
is large.  Large $M$ values are attractive for several reasons, including the fact that one may relate $M$
 to the grand unified scale. On the other hand, all $M$ values are technically natural, given that when $M$ vanishes the global symmetry structure of the Lagrangian is enhanced: $U(1)_{B-L}$ is a symmetry of the Lagrangian if $M=0$, so that $M$ is often referred to as the lepton number breaking scale. This point was recently emphasized in \cite{SeeSaw_LSND}. Recent analyses have also revealed that there are several low-energy choices for the seesaw energy scale that allow one to address outstanding problems in particle physics and astrophysics. The main reason for this is that, unlike in the high-energy seesaw, in a low-energy seesaw the mostly right-handed states do not decouple but remain as kinematically accessible sterile neutrinos.

 The data reported by the LSND short baseline neutrino oscillation experiment
\cite{LSND} can be explained by postulating the existence of light
($m\sim 1-10$~eV) sterile neutrino states.  This result is currently
being tested by the Fermilab MiniBooNE experiment \cite{minib} and, if confirmed, will lead the community to seriously contemplate the existence of light, SM singlet fermions.  It was pointed out in \cite{SeeSaw_LSND} (see also \cite{light_rhn1}) that if $M\sim 1-10$~eV, the right-handed seesaw neutrinos can easily play the role of the LSND sterile neutrinos. There is also evidence for mixed sterile neutrinos at other energy scales: eV sterile neutrinos aid in heavy element nucleosynthesis in supernovae, keV sterile neutrinos can help explain the peculiar velocity of pulsars, and remain viable warm dark matter candidates. In the past several months, it has been shown that the seesaw right-handed neutrinos may play the role of all these astrophysically/cosmologically inspired sterile neutrinos \cite{nuSM_dark,nuSM_kicks}.

In this paper, we consider the phenomenology of low-energy ($M\lesssim 1$~keV) seesaw scenarios, extending the analysis performed in \cite{SeeSaw_LSND} in several ways. In Sec.~\ref{sec:seesawMass}, we review the generation of neutrino mass via the seesaw mechanism and apply it to relatively light right-handed neutrino states. We pay special attention to the most general active--sterile seesaw mixing matrix, whose parameters are the main object of our study. In Sec.~\ref{sec:pheno}, we review the several different ``evidences'' for sterile neutrinos, and discuss whether these can be ``fit'' by the low-energy seesaw. We concentrate on exploring solutions that can accommodate at the same time the LSND anomaly and the astrophysical processes outlined above, but also discuss different combinations of the seesaw parameters capable of explaining only the astrophysics-related observables.
In Sec.~\ref{sec:future}, we examine other experimental probes that can be used to explore low-energy seesaws -- regardless of their relationship to the LSND anomaly, pulsar kicks, and warm dark  matter. We concentrate on the prospects of
current/future tritium beta-decay and neutrinoless double-beta decay
experiments. We conclude in
Sec.~\ref{sec:Conclusion} by summarizing our results, commenting
on the plausibility of this scenario, and offering a general outlook for the future.

\section{The Seesaw Mechanism and electron volt neutrino masses:
preliminaries}
\label{sec:seesawMass}

In order to account for nonzero neutrino masses, we add to the SM particle content
 three $SU(3)_c
\times SU(2)_L \times U(1)_Y$ gauge singlet fermion states $N_i$, conventionally
referred to as right-handed neutrinos. While
sterile under SM gauge interactions, right-handed neutrinos may still be charged under new,  currently unknown gauge transformations.  Such interactions, if at all present, are neglected in our analysis.

The most general renormalizable Lagrangian consistent with SM gauge invariance is
\begin{equation}
\label{eq:seesaw}
{\cal L}_{\nu}={\cal L}_{\rm old} - \lambda_{\alpha i}\bar{L}^{\alpha}HN^i-\sum_{i=1}^3\frac{M_{i}}{2}{\overline N^c}^iN^i + H.c.,
\end{equation}
where ${\cal L}_{\rm old}$ is the traditional SM Lagrangian, $H$ is the Higgs weak doublet, $L^{\alpha}$,  $\alpha=e,\mu,\tau$, are lepton weak doublets, $\lambda_{\alpha i}$ are neutrino Yukawa couplings, and $M_i$ are Majorana masses for the $N_i$. We choose, without loss of generality, the Majorana  mass matrix $M_R$ to be diagonal and its eigenvalues $M_i$ to be real and positive. We also choose the charged lepton Yukawa interactions and the charged weak current interactions diagonal so that all physical mixing elements are contained in the neutrino sector.

After electroweak symmetry breaking (when $H$ develops a vacuum expectation value $v$), ${\cal L}_{\nu}$ will describe, aside from all other SM degrees of freedom, six neutral massive Weyl fermions --- six neutrinos. The resulting mass terms can be expressed as:
\begin{equation}
\mathcal{L}_\nu \supset \frac{1}{2}\left(
                                \begin{array}{cc}
                                  \overline{\vec{\nu}} &
                                  \overline{\vec{N}^c}
                                \end{array}
                             \right)
                             \left(
                               \begin{array}{cc}
                                 0 & \mu \\
                                 \mu^T & M_R \\
                               \end{array}
                             \right)
                             \left(
                               \begin{array}{c}
                                 \vec{\nu}^c \\
                                 \vec{N} \\
                               \end{array}
                             \right)
\label{eq:lagrangian},
\end{equation}
where $\mu\equiv\lambda v$, and  $\vec{\nu}$ and $\vec{N}$ are
vectors of the three active neutrinos $(\nu_e,\nu_\mu,\nu_\tau)$ and
the three right-handed, sterile states, respectively.  Each entry in
the symmetric mass matrix of Eq~(\ref{eq:lagrangian}) is itself a
$3\times3$ matrix of mass parameters. Diagonalization of the mass
matrix yields eigenstates with Majorana masses that mix the
active--active states, related via the standard lepton mixing
matrix $V$, and the active--sterile states.  The physical neutrinos
are thus linear combinations of all active and sterile states.
Throughout, we will work in the ``seesaw limit,'' defined by $\mu
\ll M_R$.  In this case, there are three mostly active light
neutrinos and three mostly sterile heavy neutrinos where `mostly' is
determined by the induced mixing parameters.

In the seesaw limit, the diagonalization is simple. 
are real:
\begin{equation}
\left(
  \begin{array}{cc}
    0 & \mu \\
    \mu^T & M_R \\
  \end{array}
\right) = \left(
            \begin{array}{cc}
              1 & \Theta \\
              -\Theta^T & 1 \\
            \end{array}
          \right)
          \left(
            \begin{array}{cc}
              V & 0 \\
              0 & 1 \\
            \end{array}
          \right)
          \left(
            \begin{array}{cc}
              m & 0 \\
              0 & M_R \\
            \end{array}
          \right)
          \left(
            \begin{array}{cc}
              V^T & 0 \\
              0 & 1 \\
            \end{array}
          \right)
          \left(
            \begin{array}{cc}
              1 & -\Theta \\
              \Theta^T & 1 \\
            \end{array}
          \right) + \mathcal{O}(\Theta^2)
\label{eq:diag},
\end{equation}
where $m$ is the diagonal matrix of light neutrino masses and
$\Theta$ is a matrix of active--sterile mixing angles found from the
relations
\begin{eqnarray}
\Theta M_R &=& \mu, \label{eq:mu} \\
 \Theta M_R \Theta^T &=& - V m V^T \label{eq:theta}.
\end{eqnarray}
The elements of $\Theta$ are small (${\cal O}(\mu/M_R)$), and the standard seesaw relation ($VmV^T=-\mu{M}^{-1}_{R}\mu^T$)
is easily obtained by combining Eqs.~(\ref{eq:mu}) and Eqs.~(\ref{eq:theta}). On the other hand, using Eq.~(\ref{eq:theta}), we can relate the mixing parameters in
$\Theta$ to the active--active mixing angles contained in $V$ and the
neutrino mass eigenvalues.  In the case of observably light sterile
neutrino masses, as considered in our analysis, this equation is
very useful, as it places testable constraints on observable
quantities.
The general solution (first discussed in detail in \cite{Casas:2001sr})
of Eq.~(\ref{eq:theta}) is
\begin{equation}
\Theta = -Vm^{1/2}OM_R^{-1/2} \label{eq:thetaM_constraint},
\end{equation}
where $O$ is an orthogonal
matrix parameterized by three mixing angles $\phi_{12},\phi_{13},\phi_{23}$.\footnote{In general, $O$ is a \emph{complex} orthogonal matrix. Here, however, we will restrict our analysis to \emph{real} neutrino mass matrices, unless otherwise noted.} Physically,
the mixing matrix $O$ is a consequence of our freedom to choose the
form of $M_R$.  An illustrative example is found by considering the
mass ordering of $M_R$ in its diagonal form.  The reordered matrix
$M_R(m_i\leftrightarrow m_j)$ is equivalent to a $\pi/2$ rotation in
the $i-j$ plane and therefore represents the same physics as the
original matrix, as it should.  In other words, $O M_R O^T$ is the
physically relevant object, as opposed to $O$ and $M_R$ separately.
This object, when constrained to be real, has six free parameters that we shall refer to as
``heavy parameters:'' $\phi_{12},\phi_{13},\phi_{23},M_1,M_2,M_3$.

Using Eq.~(\ref{eq:thetaM_constraint}),  the $6\times6$
unitary neutrino mixing matrix is
\begin{equation}
U = \left(
 \begin{array}{cc}
   V & \Theta \\
     -\Theta^T V & 1 \\
      \end{array}
       \right)
        \label{eq:U}.
\end{equation}
Note that, up to ${\cal O}(\Theta^2)$ corrections, $V$ is unitary.
$U$ is entirely described in
terms of the three light mixing angles, six mass eigenvalues and
three angles $\phi_{ij}$.  Many combinations of these have been
measured or constrained via oscillation searches, cosmology and
astrophysics.  In particular, the two active neutrino mass-squared
differences and mixing angles have been measured
\cite{neutrino_review,global_analy,pdg}, thus leaving free the six heavy parameters and
the absolute scale of active neutrino masses.\footnote{The active neutrino mass hierarchy, normal vs. inverted,
is another (discrete) free parameter.}  With $U$, the corresponding neutrino mass
values and the SM couplings we can calculate all observable
quantities and compare them with data.

It is natural to wonder how well the seesaw approximation holds
once one starts to deal with $M_R$ values around  1~eV. From
Eq.~(\ref{eq:mu}), it is clear that one can choose for the expansion parameter  $\sqrt{m/M_R}$.  In all scenarios considered here, $\sqrt{m/M_R}<0.5$ (for $M\sim 1$~eV and $m \sim 0.3$~eV).  In the worst case scenario, therefore, first order corrections are
$55\%$ of the leading order terms, while second order
corrections are near $30\%$.   Corrections
to most observables of interest are much smaller than this because they are
suppressed by larger right-handed neutrino masses.
The first non-trivial correction to Eq.~(\ref{eq:thetaM_constraint}) occurs at second
order and we find that, for the ambitions of this paper, all approximations are under control.
This simple argument has been verified numerically for the most worrisome cases.

Before concluding this section, we wish to add that  operators that lead, after electroweak symmetry breaking,  to Majorana masses for the left--handed
neutrino states ($M_L$) are also allowed if one introduces $SU(2)_L$ Higgs boson triplets  or  nonrenormalizable operators to the SM Lagrangian.
While we neglect these ``active'' Majorana masses, we caution
the reader that the existence of such terms would alter
our results significantly. In particular, assuming the seesaw approximation holds, Eq.~(\ref{eq:theta}) would read
\begin{equation}
VmV^T+\Theta M_R \Theta^T=M_L,
\label{eq:withML}
\end{equation}
which leads to a relationship between $\Theta$, $m$, and $M_R$ different from Eq.~(\ref{eq:thetaM_constraint}). If this were the case, for example, it would no longer be true that the largest $\Theta$ value (in absolute value) is constrained to be smaller than $(m_{\rm max}/M_{\rm R,min})^{1/2}$, where $m_{\rm max}$ is the largest element of $m$, while $M_{\rm R,min}$ is the smallest element of $M_R$. On the other hand,  all objects on the left-hand side of Eq.~(\ref{eq:withML}) are observables. Hence, in the case of a low-energy seesaw, one can expect, in principle, to be able to test whether there are contributions to the neutrino mass matrix that are unrelated to the presence of right-handed neutrinos. By measuring $V$, $m$, $M_R$, and $\Theta$, one can establish whether $M_L$ is consistent with zero.

\section{Oscillation Phenomenology and Current Evidence for Low-Energy Seesaw}
\label{sec:pheno}

Here we examine a number of experimental and observational anomalies
that may be explained by light sterile neutrinos. More specifically,
we explore what these can teach us about the currently unknown
parameters of the seesaw Lagrangian, described in detail in
Sec.~\ref{sec:seesawMass}. In all cases we assume 3 mostly active
and 3 mostly sterile neutrinos and, most of the time, will
concentrate on a $3+2+1$ picture of neutrino mass eigenstates, that
is, three mostly active sub-eV neutrinos, two mostly sterile eV
neutrinos and one almost completely sterile keV neutrino. The hope
is that the heavier state can  account for warm dark matter (section
\ref{subsec:WDM}) or pulsar kicks (Sec.~\ref{subsec:Kicks}), which
both require at least one keV neutrino, while the other two mostly
sterile states help ``explain'' the existing oscillation data where,
for all practical purposes, the heaviest neutrino decouples and we
are left with an effective $3+1$ or $3+2$ picture.  We remind
readers that a third possibility (2+2\footnote{It would have been
rather difficult to construct a 2+2 neutrino mass hierarchy using
the seesaw Lagrangian.}) is currently ruled out by solar and
atmospheric data \cite{CPTVorSterile,rule_out_4nu,Maltoni:2003yr}
and will be ignored. $3+1$ schemes that address the LSND anomaly are
also disfavored by global analysis of short baseline oscillation
experiments
\cite{CPTVorSterile,rule_out_4nu,Maltoni:2003yr,short_bl_analysis},
and for this reason, we mostly concentrate on $3+2$ fits to the LSND
anomaly \cite{short_bl_analysis}.

Our analysis method is as follows:  For each experimental probe
considered we perform a $\chi^2$ ``fit'' of the mixing matrix $U$,
given by Eq.~(\ref{eq:U}), and neutrino masses to the ``data'', and
extract the region of parameter space that best explains the data.
In most cases, we allow the light mixing angles and mass squared
differences to vary within their $1\sigma$ limits (according to
\cite{global_analy}),\footnote{In the case of 3+1 ``fits'' ({\it
cf.} Eqs.~(\ref{U_31_i},\ref{U31_n})), we kept the active  neutrino
parameters fixed at their best-fit values.} the angles $\phi_i$ to
vary unconstrained within their physical limits of $0-2\pi$, and the
lightest active neutrino mass eigenvalue $m_l$ to vary unconstrained
between $0-0.5$ eV. The quotation marks around  ``fit'' and ``data''
are meant to indicate that our methods are crude, in the sense that
we are fitting to previously processed experimental data, assuming a
diagonal correlation matrix with Gaussian uncertainties. In order to
avoid the subtleties involved in such a ``fit to a fit'', we
hesitate to mention actual confidence intervals, but are compelled
to do so for lack of a better measure of an allowed region.  We
sometimes present our best fit parameter points along with
confidence intervals, but warn the reader to avoid strict
interpretations of these numbers. While crude, our methodology of
error analysis and fitting provides a very useful  instrument for
identifying whether (and how) the low-energy seesaw can accommodate
a particular combination of data sets.

Before proceeding, it is useful to cement our notation. The neutrino masses will be ordered in ascending order of magnitude from $m_1$ to $m_6$ in the case of a normal active neutrino mass hierarchy ($m_2^2-m_1^2<m_3^2-m_1^2$), while in the case of an inverted mass hierarchy they are ordered $m_3<m_1<m_2<m_4<m_5<m_6$ (in this case $|m_3^2-m_1^2|>m_2^2-m_1^2$). The states with masses $m_{1,2,3}$ are mostly active, while those with masses $m_{4,5,6}$ are mostly sterile. Elements of the mixing matrix are referred to as $U_{\alpha i}$, where $\alpha=e,\mu,\tau,s_1,s_2,s_3$ ($s$'s are the right-handed neutrino degrees of freedom) and $i=1,2,3,4,5,6$. We also define $\Delta m^2_{ji}=m_j^2-m_i^2$ and will refer to the lightest active neutrino mass as $m_l$. In the case of normal (inverted) active neutrino mass hierarchy $m_l=m_1$ ($m_l=m_3$).

\subsection{Short baseline oscillation constraints}
\label{subsec:LSND}

Here we analyze the constraints imposed on the unknown mixing
parameters by current neutrino oscillation data. We will assume that
all solar, reactor, long-baseline and atmospheric data are properly
fit with active--active oscillations, and that constraints on the
other seesaw parameters will be provided mostly by short-baseline
accelerator experiments. It is interesting to note that the
inclusion of the angles $\phi_{ij}$ introduces enormous freedom into
the system. Any one active--sterile mixing angle contained in
$\Theta$ can always be set to zero by an appropriate choice of $O$.
In fact, all but three elements may be set to zero simultaneously,
with only a single non-zero element in each row and column. In this
case, these are constrained to be around $\sqrt{m_l/M_{Ri}}$ where
$i$ is the column of the non-zero element. This is especially true
when the mostly active neutrino masses are quasi-degenerate. An
important ``sum rule of thumb'' is the following. For a given
right-handed neutrino mass $M_i$, the active--sterile mixing angle
squared is of order $m/M_i$, where $m$ is a typical active neutrino
mass. One can always choose parameters so that, for at most, two
values of $\alpha=e,\mu,\tau$, $U_{\alpha i}$ are abnormally small.
In that case, however, the ``other'' $U_{\alpha i}$ is constrained
to saturate the bound $|U_{\alpha i}|^2\lesssim m_l/M_i$.

The most compelling evidence for light sterile neutrinos comes from
the short baseline oscillation experiment by the Liquid Scintillator
Neutrino Detector (LSND) collaboration at Los Alamos. Using a $\sim
30$ MeV $\overline{\nu}_\mu$ beam they observed a better than
$3\sigma$ excess of $\overline{\nu}_e$--like events above their
expected background at their detector some 30~m away
 from the production point \cite{LSND}. This evidence of $\overline{\nu}_\mu \leftrightarrow
\overline{\nu}_e$ oscillation requires a mass-squared difference
greater than $1~{\rm eV}^2$, clearly incompatible with the small
mass-squared differences observed between the active neutrinos.
Several mechanisms, such as CPT-violation \cite{CPTV,CPTVorSterile},
Lorentz invariance  violation \cite{lorentz_v}, quantum decoherence
\cite{decoherence}, sterile neutrino decay \cite{sterile_decay} and,
of course, oscillation into sterile neutrinos have been proposed to
explain this result.  Here we concentrate on the last possibility.

In order to take into account all short baseline data we ``fit'' our
mixing parameters and masses to the results of the $3+2$ performed in
\cite{short_bl_analysis}, which are summarized in Table
\ref{tab:LSND_fit} \cite{3p2fit}.  Here we assume that the heaviest, mostly sterile state does
not participate effectively in LSND oscillations. This is guaranteed to happen if $m_6\gtrsim 10$~eV.
 On the other hand, $|U_{\alpha 6}|^2$ are partially constrained by our attempts to accommodate LSND data
 with seesaw sterile neutrinos, as will become clear in the next subsections.
\begin{table}[b]
\centering \caption{Parameter values used in our analysis.  These
were extracted from a fit to all short baseline neutrino oscillation
experiments including LSND within the $3+2$ scenario
\cite{short_bl_analysis,3p2fit}. $1\sigma$ indicates a rough estimate of the 1~sigma allowed range for the different parameters.}
\begin{tabular}{c|c c c c c c}
   & $U_{e4}$ & $U_{\mu 4}$ & $U_{e5}$ & $U_{e\mu 5}$ & $\Delta m^2_{41}~(\rm{eV}^2)$ & $\Delta m^2_{51}~(\rm{eV}^2)$ \\
   \hline
  Central Value & 0.121 & 0.204 & 0.036 & 0.224 & 0.92 & 22 \\
  $1\sigma$ & 0.015 & 0.027 & 0.034 & 0.018 & 0.08 & 2.4 \\
\end{tabular}
\label{tab:LSND_fit}
\end{table}

\begin{figure}
\begin{center}
\includegraphics[scale=0.6]{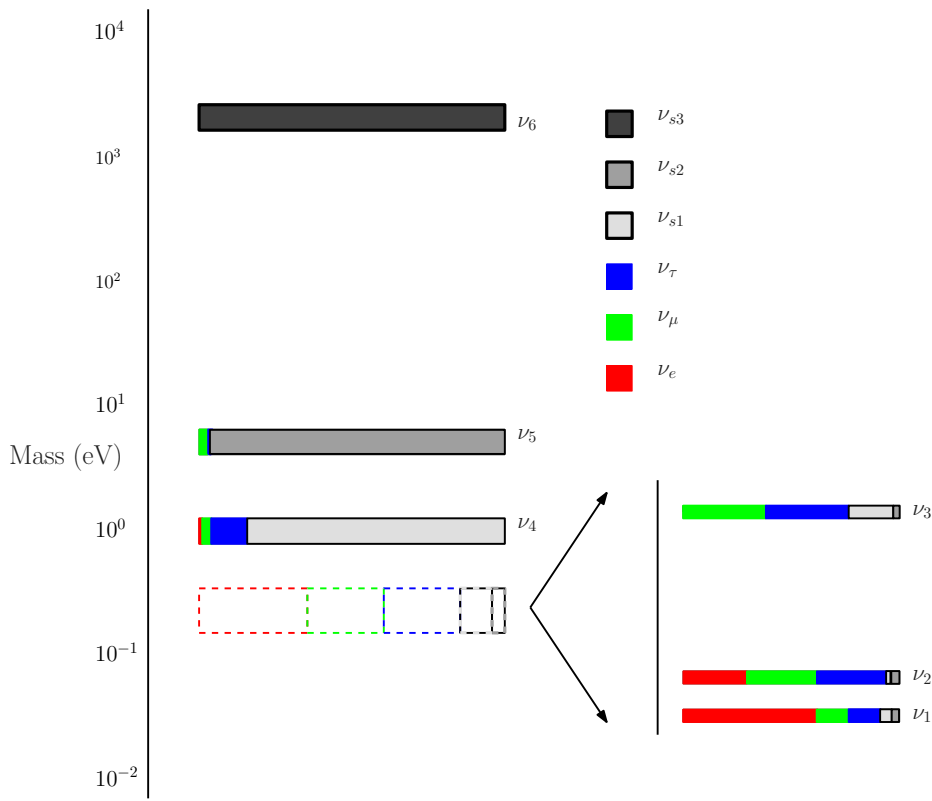}
\caption{Neutrino mass eigenstate spectrum, along with the flavor
composition of each state.  This case accommodates all neutrino
oscillation data, constraints from r-process nucleosynthesis in
supernovae, and may help explain anomalous pulsar kicks (see text
for details).  While we choose to depict a normal hierarchy for the
active neutrino states, an inverted active neutrino mass hierarchy
would have yielded exactly the same physics (as far as the
observables considered are concerned). } \label{fig:spectrum}
\end{center}
\end{figure}

 We find that  $m_l$, the
lightest neutrino mass, is constrained to lie between
$(0.22-0.37)~\rm{eV}$, with a ``best fit'' value of $0.29~\rm{eV}$.
Thus, the active neutrino mass spectrum is predicted to be
quasi-degenerate.  A sample $6\times 6$ neutrino mixing matrix that
fits all data is
\begin{equation}
U_{3+2} = \left(
  \begin{array}{cccccc}
   0.8301 &   0.5571 &   0.001365 &   0.1193 &  -0.009399 &
   -0.006513\\
  -0.3946 &   0.5866 &   0.7072 &   0.2016 &   0.2262 &
  0.0003363\\
   0.3932 &  -0.5879 &   0.7070 &   0.4760 &  -0.0949 &   0.001470\\
  -0.2067 &   0.09514 &  -0.4792 &   1 &                  0&
  0\\
   0.1343 &  -0.1832 &  -0.09284 &                  0&   1&
   0\\
   0.004963 &  0.004295 &  -0.001268 &                  0 &                 0&
   1\\
  \end{array}
\right),
\label{U_32}
\end{equation}
while the associated masses are $m_1\simeq m_2\simeq m_3=0.28$~eV, $m_4=1.0$~eV, $m_5=4.7$~eV, and $m_6=6.4$~keV. Note that the matrix in Eq.~(\ref{U_32}) is only approximately unitary, up to corrections of order 25\%. This result agrees qualitative with those obtained in \cite{SeeSaw_LSND}. The neutrino masses and mixings obtained in this ``fit'' are depicted in Fig.~\ref{fig:spectrum}. We will use the results of ``fits'' similar to this one throughout the paper.

One can also  aim at a (currently disfavored)
$3+1$ LSND fit.\footnote{This is easily accomplished by requiring $m_5\gtrsim 10$~eV.}. In this case,  much lower $m_l$ values are also
allowed, extending well into the hierarchical spectrum range. In
this case, all $m_l$ values above $0.01~\rm{eV}$ and $0.03~\rm{eV}$ are
allowed, assuming an inverted and normal mass hierarchy,
respectively. This is to be compared with the results found in \cite{SeeSaw_LSND}, where only trivial choices for $O$ were considered. Examples that ``fit'' all oscillation data include, for an inverted active mass hierarchy:
$m_1\simeq m_2=0.066$~eV, $m_3=0.043$~eV, $m_4=0.96$~eV, $m_5=5$~keV, and $m_6=10$~GeV, together with
\begin{equation}
U_{3+1}^{\rm inverted} = \left(
  \begin{array}{cccccc}
   0.8305 &  0.5571 &  0 &  0.1359 &  -0.00009142 &  -0.000002198 \\
  -0.3939 &   0.5872 &   0.7071 &   0.2046 &   0.00005000 &
  0.000001202\\
   0.3939 &  -0.5872 &   0.7071 &  -0.04421 &  -0.003236 &
   -0.0000001857\\
  -0.01486 &  -0.2218 &  -0.1134 &   1 &                  0 &
  0\\
   0.001370 &  -0.001878 &   0.002253 &                  0 &  1         &
   0\\
   0.000002372 &   0.0000004094 &  -0.0000007187    &               0    &              0 &
   1\\
  \end{array}
\right).
\label{U_31_i}
\end{equation}
For a normal mass hierarchy, we find that $m_1=0.055$~eV, $m_2=0.056$~eV , $m_3=0.0744$~eV, $m_4=0.96$~eV, $m_5=5$~keV, $m_6=10$~GeV, and
\begin{equation}
U^{\rm normal}_{3+1} = \left(
  \begin{array}{cccccc}
   0.8305 &   0.5571 &                  0 &  0.1173 &  -0.002100 &
   0.000001418\\
  -0.3939 &   0.5872 &   0.7071 &   0.2176 &   0.0004625 &
  -0.000001364\\
   0.3939 &  -0.5872 &   0.7071 &   0.09802 &   0.002804 &
   0.000001283\\
  -0.05028 &  -0.1355 &  -0.2231 &   1 &                  0 &
  0\\
   0.0008214 &   0.002545 &  -0.002310 &                  0  & 1       &
   0\\
  -0.000002220 &   0.0000007646 &   0.00000005663 &                  0  &                0 &
  1\\
  \end{array}
\right),
\label{U31_n}
\end{equation}
``fit'' all oscillation data quite well.

Note that a null result from MiniBooNE is bound to place significant limits on the seesaw energy scale. If all right-handed neutrino masses are similar, the effective mixing angle that governs $\nu_\mu\rightarrow\nu_e$ transitions is $\sin^22\theta_{\rm MiniBooNE}\lesssim 4 m^2/M^2$. Hence, a null result at LSND would rule out a seesaw energy scale $M$ lighter than 6~eV, assuming all active neutrino masses $m$ are around 0.1~eV \cite{minib}. This limit is sensitive to the lightest neutrino mass $m_l$ and can be somewhat relaxed (similar to how we obtain a good 3+2 to all neutrino data) by postulating a (mild) hierarchy of right-handed neutrino masses and by assuming that sterile-electron and sterile-muon neutrino mixing is suppressed with respect to naive expectations for the lightest mostly sterile state(s). For larger values of $m_l$, $M$ values around 10~eV are already constrained by $\nu_{\mu}\to\nu_e$ searches at the NuTeV \cite{nutev} and NOMAD \cite{nomad} experiments, and $\nu_{\mu}\to\nu_{\tau}$ searches at CHORUS \cite{chorus}.

\setcounter{footnote}{0}
\subsection{Cosmological and Astrophysical Constraints, Warm Dark Matter}
\label{subsec:WDM}

Very light sterile neutrinos that mix with the active neutrinos are
constrained by several cosmological and astrophysical observables.
The ``seesaw'' right-handed neutrinos are no exception. Given that
active--sterile mixing angles $|U_{\alpha i}|^2\lesssim m_l/m_i$
($\alpha=e,\mu,\tau$, $i=4,5,6$), it turns out that for ``standard
cosmology,'' the right-handed neutrinos thermalize with the early
universe thermal bath of SM particles, as long as the reheat
temperature is higher than their Majorana masses. For the low seesaw
energy scales we are interested in, this is a problem. For the
values of $M_R$ under consideration here, thermal right-handed
neutrinos easily overclose the universe. Smaller $m_l$ values
($m_l\lesssim 10^{-5}$~eV) lead to the possibility that right-handed
neutrinos are the dark matter, as recently discussed in the
literature \cite{nuSM_dark,nuSM_kicks}. We comment on this and other
possibilities shortly.

\begin{figure}
\begin{center}
\includegraphics[angle = 270,scale=0.60]{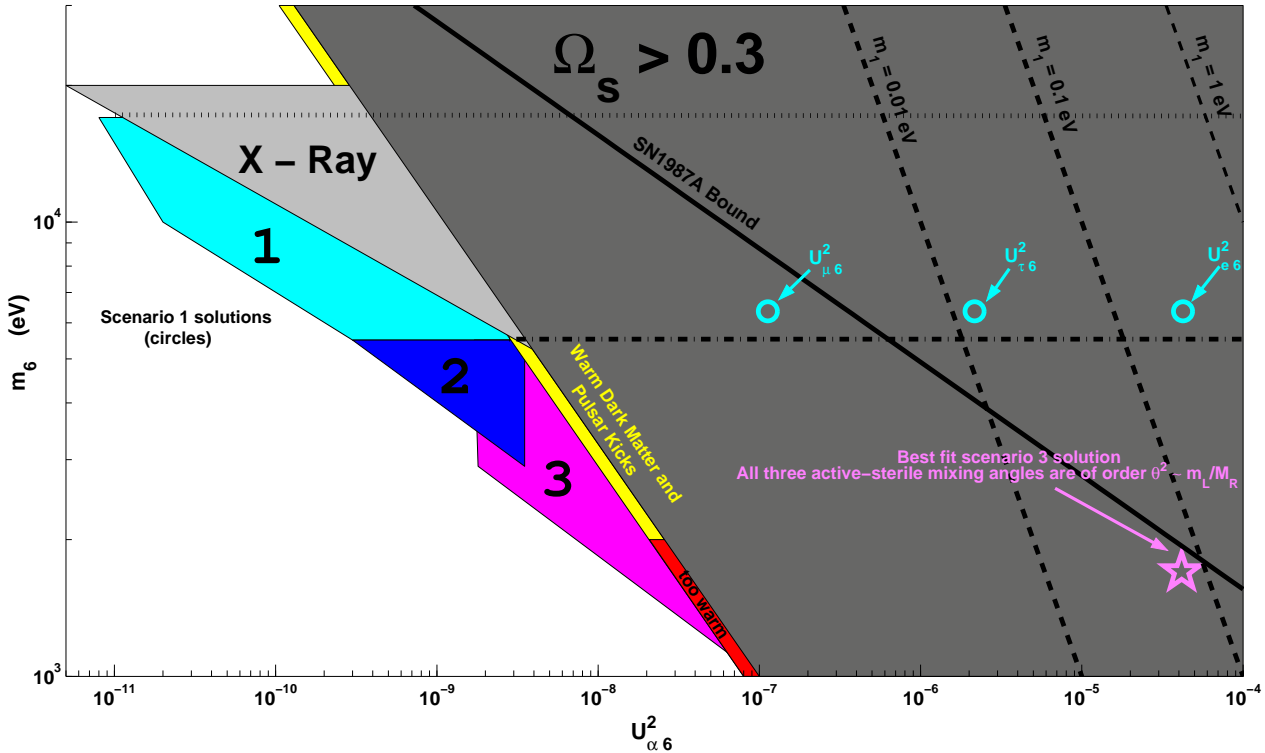}
\caption{Adapted from \cite{kicks_gen}.  Cosmological and
astrophysical constraints on the $|U_{\alpha 6}|^2\times m_6$-plane.
In the large dark grey region, the density of a thermal $\nu_6$
population is $\Omega_s>0.3$, while the light grey `X-ray' region is
disfavored by X-ray observations.  The regions labeled 1,2,3 are
preferred if one is to explain anomalous pulsar kicks with
active--sterile oscillations inside supernovae. Regions 1 and 3
qualitatively extend inside the  $\Omega_s>0.3$ part of the plane as
indicated by the horizontal dotted and dash-dotted lines,
respectively. The regions `Warm Dark Matter'' and ``Too Warm Dark
Matter'' are meant to represent the region of parameter space where
thermal $\nu_6$ qualifies as a good (or bad) warm dark matter
candidate. The region above the solid diagonal line is disfavored by
the observation of electron (anti)neutrinos from SN1987A. The
diagonal dashed lines correspond to $U_{e6}^2=m_l/m_6$, for
different values of $m_l$.
 Also shown is our ``best fit'' sterile solution
for different  pulsar kick scenarios, assuming the $3+2$ LSND fit
for the lighter states.  The regions one and three best fit values
are represented by circles and a star, respectively. See text for
details.} \label{fig:kicks}
\end{center}
\end{figure}

Fig.~\ref{fig:kicks} depicts the region of the $|U_{e6}|^2 \times
m_6$-plane in which the contribution of the heaviest neutrino
$\nu_6$ to $\Omega$ (the normalized energy density of the universe,
$\rho/\rho_c$) is larger than 0.3 (dark region). The same constraint
roughly applies for all $\alpha=e,\mu,\tau$ and $i=4,5,6$. The
dashed diagonal lines correspond to $|U_{\alpha 6}|^2=m_l/m_6$, for
different values of $m_l$. All lines lie deep within the dark
$\Omega_s>0.3$ region.

For smaller values of $M_R$, the situation is also constrained. For
$M_R$ values below tens of eV, thermal sterile neutrinos contribute
to the amount of hot dark matter in the universe
\cite{wmap,seljak,Hannestad:2006mi}. Right-handed neutrinos will
thermalize as long as $m_i\sin^2\theta_{i\alpha}\gtrsim 5\times
10^{-4}$~eV \cite{thermalize,smirnov_renata}. In low energy seesaws,
this roughly translates into $m\gtrsim 10^{-3}$~eV, where $m$ is the
active neutrino mass scale. For the cases of interest here,
$m\gtrsim \sqrt{\Delta m^2_{21}}\sim 10^{-2}$~eV, in which case
$m_i$ values above 0.6~eV are ruled out (according to
\cite{Hannestad:2006mi}).\footnote{In our ``best fit'' 3+2 solution
to the LSND data, the sum of all active neutrino masses violates
slightly the constraint obtained in  \cite{Hannestad:2006mi}. This
problem can be easily evaded if we choose $m_l$ values close to the
lower end of the ``allowed region''. Other ``mild'' non-standard
cosmology effects \cite{Beacom:2004yd} are also known to alleviate
the hot dark matter bound on neutrino masses.} Note that any sterile
neutrino solution to the LSND anomaly faces a similar problem, which
must be resolved with non-standard cosmology.

One way to avoid the bounds described above (see also, for example,
\cite{Abazajian:2004aj}) is to consider that the reheating
temperature of the universe is very low, say $T_r\lesssim 5$~MeV
\cite{low_reheat}. This way, right-handed neutrinos, in spite of
their ``large'' mixing angles, never reach thermal equilibrium in
the early universe and neither overclose the universe, nor
contribute to the amount of hot dark matter. Unless otherwise noted,
this is the assumption we make here.

Under these circumstances, it is interesting to consider whether
light seesaw right-handed neutrinos still qualify as good warm dark
matter. This could happen if their production in the early universe
was non-thermal.  One concern surrounding warm dark matter is
whether it is ruled out by large scale structure surveys. Here, we
will not add to this discussion but refer readers to the recent
literature on the subject \cite{warm_recent}. A brief summary of the
situation is as follows: constraints on warm dark matter can be
translated into a lower bound on the mostly sterile neutrino mass.
The lower bound has been computed by different groups, and lies
somewhere between 3 and 14~keV \cite{warm_recent}. Different lower
bounds depend on several issues, including which subset of Lyman
alpha-forest data is taken into account.

Another constraint on potential dark matter sterile neutrinos comes
from the observation of X-rays originating in galactic clusters.
Such regions of the universe should be overdense with warm dark
matter heavy neutrinos, which can be directly observed via their
radiative decay $\nu_6 \rightarrow \nu_i + \gamma$ \cite{Xray}.
Bounds from X-ray observations have been summarized very recently in
\cite{Kusenko:2006wa}. Combining the results of
\cite{Kusenko:2006wa} and Fig.~\ref{fig:kicks}, we find that for
lightest neutrino masses larger than $10^{-2}$~eV, such bounds can
only be avoided for $m_i\lesssim 100$~eV, where large scale
structure constraints on warm/hot dark matter are severe. This
qualitative analysis indicates that seesaw sterile neutrinos cannot
simultaneously fit the LSND data and serve as cold dark matter.

On the astrophysics side, the most severe constraint on light,
sterile neutrinos is provided by the observation of electron
(anti)neutrinos coming from SN1987A. The current analysis consists
of comparing the model-dependent neutrino flux at the surface of the
neutrinosphere with that detected on Earth.  Large sterile neutrino
mixing and mass would result in modification/depletion of the
detected neutrino signal (for a recent detailed discussion, see
\cite{sterile_probe}). Although only twenty neutrinos were observed
in this event, one can still place bounds on sterile-active neutrino
mixing. As far as ``LSND'' sterile neutrinos are concerned, these
bounds are still weaker than those obtained by the null short
baseline oscillation experiments \cite{SN1987_bound} and therefore
already accounted for in our analysis.  Heavier right-handed
neutrinos can, however, be excluded by SN1987A neutrino data.
Fig.~\ref{fig:kicks} depicts the region of parameter space excluded
by SN1987A data (region to the right of solid, diagonal line). This
bound is defined by $m_i\sqrt{2}\sqrt{U_{\alpha i}}>$0.22~keV
\cite{sn_bound,low_reheat}. See also \cite{smirnov_renata}.
According to Fig.~\ref{fig:kicks}, supernova bounds force the seesaw
scale to be below a few keV for $m_l$ values above $0.01$~eV.


\subsection{Pulsar Kicks}
\label{subsec:Kicks}

 Pulsars are born from the
gravitational collapse of the iron core of a massive star.  These
core collapse supernova are an excellent source of neutrinos,
producing all (active) flavors copiously (see \cite{SN_review} for a
detailed review). Current observations point to the fact that some
pulsars move with peculiar velocities much greater than those
expected from an asymmetric supernova explosion mechanism.
Quantitatively, current three dimensional models yield velocities up
to $200~\rm{km/s}$ \cite{pulsar_disp}, while pulsars moving at
speeds as high as $1600~\rm{km/s}$ have been observed.

Since roughly $99\%$ of the approximately $10^{53}~\rm{ergs}$ of
energy released in a core collapse supernova is in the form of
neutrinos, it is reasonable that neutrino physics provides a
solution to this anomaly. At these rates a small $(1-3) \%$
asymmetry in neutrino emission can account for the observed large
pulsar velocities.  Neutrinos are always \emph{produced}
asymmetrically in the polarized medium of the proto-neutron star,
due to the left--handed nature of their interactions. Unfortunately,
asymmetric production cannot solve this problem because the
associated medium densities are such that neutrinos undergo multiple
scattering within the star's interior, eventually diffusing out of
an effective surface, called the neutrinosphere, with all initial
asymmetries washed away. Several distinct mechanisms have been
formulated to sidestep this fact. Specifically, the existence of
large neutrino magnetic moments has been explored in
\cite{mmoment_kick}, and can be tested in next generation neutrino
scattering experiments \cite{nu_e_scattering,munu,texono}.  Proposed
solutions also exist which utilize standard three flavor neutrino
oscillations, where $\nu_\mu$ and $\nu_\tau$, appearing between
their neutrinosphere and the larger $\nu_e$ neutrinosphere, can
stream unhindered out of the star \cite{active_kicks}.  This
solution is, however, currently disfavored by terrestrial
oscillation experiments.

We concentrate on  the case of oscillations into sterile neutrinos,
which can proceed in various ways, depending on the mass and
coupling of the relevant neutrinos as well as the properties of the
collapsing star, including its density and magnetic field. Following
\cite{kicks_gen}, we separate and analyze these within three
distinct categories.  Each one requires the existence of a keV-scale
sterile neutrino with very small couplings to the active flavors, of
the order $10^{-4} - 10^{-5}$, especially if light sterile neutrinos
are thermally produced in the early universe. Under these
circumstances, if seesaw neutrinos are to play the role of the
sterile neutrinos responsible for pulsar kicks,  $|U_{\alpha
i}|^2\lesssim m/m_i$ ($i=4,5,6$) must lie in the $10^{-9}$ range for
$m_i\sim 10^{4}$~eV. This implies $m\sim 10^{-5}$~eV and is only
compatible with a hierarchical active neutrino mass spectrum and
very light $m_l$, as identified in \cite{nuSM_dark,nuSM_kicks}.

Here, instead, we will concentrate on identifying solutions that
will address pulsar kicks and the LSND anomaly. According to the
discussion in the previous subsection, the mostly sterile neutrino
masses $m_4$ and $m_5$ are constrained to be less than 10~eV so that
a 3+2 solution to the LSND anomaly can be obtained from the seesaw
Lagrangian. The heaviest neutrino mass $m_6$ is unconstrained, so we
are free to vary it as needed in order to attack the pulsar peculiar
velocity issue.\footnote{We can neglect the lighter sterile
neutrinos ($\nu_4$ and $\nu_5$), as they should not alter the
kicking mechanism significantly due to their small mass and
non-resonant production.} Naively, the  fraction of $\nu_{\alpha}$
($\alpha=e,\mu,\tau$) in $\nu_6$ is expected to be of the order
$U_{\alpha 6} \sim \sqrt{0.3~\rm{eV}/3\times 10^3~\rm{eV}} =
10^{-2}$, much too large to satisfy the pulsar kick plus cosmology
constraints summarized in Fig.~\ref{fig:kicks}. On the other hand,
once the $\Omega_S<0.3$ constraint is removed, the `pulsar kicks'
allowed region of the plane is significantly enlarged, as
qualitatively  indicated by the horizontal lines in
Fig.~\ref{fig:kicks}. In this case, which we must consider anyway if
we are to have agreement between LSND and searches for hot dark
matter, one can envision explaining pulsar kicks and the LSND data
simultaneously.

In scenario 1, the pulsar kick is produced via an active--sterile
MSW resonance in the core of the proto-neutron star at large
densities, greater than $10^{14}~\rm{g/cm^3}$, and magnetic fields,
near $10^{16}~\rm{G}$ \cite{kicks_res}. The effective neutrino
matter potential in material polarized by a strong magnetic field
contains a term proportional to $\vec{k} \cdot \vec{B}/|\vec{k}|$
\cite{nu_Bfield1,nu_Bfield2}, where $\vec{k}$ is the neutrino's
three-momentum and $\vec{B}$ is the local magnetic field vector.
Clearly the MSW resonance occurs at a radius that depends on
$\vec{k} \cdot \vec{B}/|\vec{k}|$, the relative orientation of the
neutrino momentum and magnetic field.  Sterile neutrinos produced at
smaller radii (higher temperatures) carry greater average momentum
than those produced at larger radii (lower temperatures), yielding
an asymmetric momentum distribution of emitted neutrinos. This
asymmetry is capable of producing the observed pulsar kicks, in the
direction of the magnetic field, when the mass and coupling of the
sterile state is near $8~\rm{keV}$ and above $1.5\times 10^{-5}$,
respectively \cite{kicks_gen}. We found the  ``best fit'' to the
LSND data (using $\nu_4$ and $\nu_5$) and pulsar kicks (using
$\nu_6$) and $m_6>5$~keV. The $|U_{\alpha 6}|^2$ and $m_6$ ``best
fit'' values are depicted in Fig.~\ref{fig:kicks}. This solution is
strongly disfavored by the observation of neutrinos from SN1987A.
The fact that $|U_{e6}|$ is much larger than the  other two
active--sterile mixing angles is due to the fact that $|U_{e4}|$ and
$|U_{e5}|$ are constrained by LSND data to be much smaller than
naive expectations (see Eq.~(\ref{U_32})). In order to reduce
$|U_{e6}|$, one would have to either reduce $m_l$ by an order of
magnitude -- which renders the 3+2 fit to oscillation data very poor
-- or increase $m_6$, which would only push  $|U_{e6}|$ deeper into
the region of parameter space ruled out by SN1987A. One can,
however, find 3+1 solutions to LSND data where $\nu_5$ could pose as
the sterile neutrino that explains why pulsar peculiar velocities
are so large (see Eqs.~(\ref{U_31_i},\ref{U31_n})).

Scenario 2 also relies on a direction-dependent MSW resonance, this
time occurring outside the core where the matter density and
temperature are much lower.  Here, both the active and sterile
neutrinos are free to stream out of the star.  The departing active
flavors still have a small interaction cross-section, $\sigma \sim
G_F^2 E_\nu^2$, and can therefore deposit energy and momentum into
the star's gravitationally bound envelope proportional to the matter
it transverses.  Via the direction-dependent resonance, neutrinos
moving in the direction of the magnetic field remain active longer,
deposit more momentum, and thus, kick the star forward.  The
observations can be explained in this case with a smaller sterile
neutrino mass and larger active--sterile coupling near $4~\rm{keV}$
and $4.5\times 10^{-5}$, respectively \cite{kicks_gen}.  In the case
of our LSND ``fit'' the data, we can constrain one of $|U_{\mu
6}|^2$ or $|U_{\tau 6}|^2$ to lie inside region 2. The other
$|U_{\alpha6}|^2$ ($\alpha=e,\tau$ or $e,\mu$), however, are
constrained to be large, and lie inside region 3.

Scenario 3 proceeds through off-resonance production of the sterile
neutrino in the proto-neutron star core \cite{kicks_nonres}.  The
amplitude for sterile neutrino production by a weak process is
proportional to $U_{\alpha 6}^m$, the effective mixing angle between
the heavy mass eigenstate and the flavor eigenstate. Initially, this
quantity is very small due to matter effects in the dense core. The
effective potential in the star's interior is quickly driven to zero
in the presence of sterile neutrino production by a negative
feedback mechanism.  If this occurs in a time less than the
diffusion time-scale for the active neutrinos, approximately
$(3-10)$~s, the mixing angle will reach its vacuum value
\cite{sterile_hotwarmcold}.  The sterile neutrinos will then be
produced and emitted asymmetrically and thus, kick the pulsar to
large velocities.  Lower limits on the vacuum mass and mixing values
are derived by requiring that the off-resonance time scale
(inversely proportional to $m_6^4 \sin^2 2\theta_{\alpha 6}$) for
the evolution of the matter potential to zero be less than about ten
seconds. This places the sterile mass and mixing at approximately
$1~\rm{keV}$ and above $5\times 10^{-5}$, respectively. Since all
three active flavors are present in equal abundances and all
contribute to the effective matter potential, the mixing angle in
question is not any particular $U_{\alpha 6}$. Rather, it is the
angle $\theta_6$ associated with the projection of $\nu_6$ onto the
space spanned by $\nu_e$, $\nu_\mu$ and $\nu_\tau$, that is
$\theta_6^2 \equiv U_{e 6}^2 + U_{\mu 6}^2 + U_{\tau 6}^2$.  From
Eq.~(\ref{eq:thetaM_constraint}) we see that $\theta_6 =
\sqrt{m_l/m_6}$ up to corrections due to the non-unitarity of $V$
and active neutrino mass differences. This is independent of  mixing
angles, and therefore cannot be tuned to be small. Our ``best fit''
region-3 solution is depicted in Fig.~\ref{fig:kicks} by a star. It
turns out that $U_{\alpha 6}$ have very similar values for
$\alpha=e,\mu,\tau$. In order to evade the SN1987A constraint, we
were forced to pick $m_l$ values close to lower bound of our 3+2
LSND ``fit'' ($m_l=0.22$~eV), so that $|U_{e4}|$, $|U_{\mu4}|$, and
$|U_{\mu5}|$ are close to the low-end of the allowed range in
Table~\ref{tab:LSND_fit}.

\subsection{Supernova Nucleosynthesis}
\label{subsec:Nucleosynthesis}

Core collapse supernova are believed to produce the observed heavy element ($A \geq 100$)
abundance through the r-process, or rapid neutron capture process.
Here we briefly review this mechanism (see \cite{rprocess_nuwind}
for a comprehensive review), as well as its facilitation by the
addition of active--sterile neutrino oscillations
\cite{nucleosynthesis1}.  This scenario begins in the neutrino
driven wind; that is, the wind of ejected nucleons driven by
neutrinos radiated from the cooling proto-neutron star. The
maintenance of equilibrium among neutrons, protons, and electron
(anti)neutrinos in neutrino capture processes leads to a
neutrino--rich environment. As the wind propagates, it cools enough
for all free protons to bind into $\alpha$ particles. In the ideal
r-process picture, as the wind cools further these $\alpha$
particles bind into intermediate size seed nuclei which later
undergo neutron capture to form the observed heavy r-process
elements.

This ideal scenario is dampened by the large number of electron
neutrinos present at the stage of $\alpha$ particle formation. These
will capture on the free neutrons, converting them to protons, which
in turn will fuse to make more $\alpha$ particles.  The end result
is a very small free neutron to $\alpha$ particle ratio, conditions
unfavorable for r-process element formation.  This is known as the
$\alpha$ effect and must be circumvented to produce the correct
distribution of heavy elements. A clear solution to this problem is
to reduce the number of electron neutrinos present at this stage,
which can be accomplished by resonant $\nu_e \rightarrow \nu_s$
conversion\footnote{In this mechanism the effective matter
potential, which depends on the number of electrons, positrons and
neutrinos, varies wildly as a function of distance from the core.
Along this radial direction there are three relevant MSW resonant
conversions that must be tracked and understood:  $\nu_e \rightarrow
\nu_s$, $\bar{\nu}_e \rightarrow \bar{\nu}_s$ and $\bar{\nu}_s
\rightarrow \bar{\nu}_e$. See \cite{nucleosynthesis1} for more
information. } \cite{nucleosynthesis1}.

The sterile neutrino solution to the r-process mechanism is modeled
and fit to the data in reference \cite{nucleosynthesis2} including
the effects of relevant nuclear physics and additional neutrino
oscillations in the star's envelope.  The analysis is expanded in
\cite{nucleosynthesis_fisscy} with the inclusion of fission cycling
of the produced heavy elements.  The analysis indicates the need for
an eV-scale sterile neutrino with an allowed parameter space much
larger than that constrained by LSND.  By itself, the requirement of successful r-process
in supernovae only weakly constrains the light
neutrino mass scale to be greater than $10^{-2}~\rm{eV}$ and
$10^{-3}~\rm{eV}$ for a $1~\rm{eV}$ and $10~\rm{eV}$ sterile
neutrino, respectively.  With regard to the LSND results, it has
been demonstrated that the $3+1$ oscillation scenario fits within
this parameter space \cite{nucleosynthesis_fisscy}. Considering that
the best fit mass-squared difference and mixing angles for the
fourth mass eigenstate, which makes up most of the lightest sterile
neutrino, is very similar between the $3+1$ and $3+2$ case
\cite{3p2fit}, it is reasonable to conclude that $\nu_e
\leftrightarrow \nu_{s4}$ resonant conversion will also fit within
this scenario.  Even oscillations into the heavier $\nu_{s5}$ state
can potentially solve this anomaly if the neutrino driven wind
expansion time-scale is sufficiently small, $ \leq 0.1~\rm{sec}$.  To
conclude this section we note that, although the sterile neutrino
solution to the supernova nucleosynthesis problem fits well within
our seesaw framework, it adds no additional constraints, and
therefore does not increase the predictability of our scenario.

\section{Other Probes of the Seesaw Energy Scale}
\label{sec:future}

Here we survey other existing and future probes of light
sterile neutrinos.  As opposed to the previous cases, these probes
are perfectly consistent by themselves.  That is, extra heavy
neutrinos are not required to solve problems within the system.
However, their addition can lead to large modifications to the
outcome of such experiments, thus rendering the eV-scale seesaw
scenario testable.  Specifically, we consider bounds from
tritium beta-decay and neutrinoless double-beta decay.  Observations in all of
these areas have already yielded useful constraints on sterile
neutrinos, and the situation is expected to improve in the next few
years.

\subsection{Tritium Beta-Decay}

The endpoint of the electron energy spectrum in the beta-decay of
tritium is a powerful probe of nonzero neutrino masses.  This
results from the decay kinematics of the system, which is
necessarily modified by the presence of a massive neutrino. The
nonzero neutrino mass effect can be understood almost entirely from
the analysis of the phase space distribution of the emitted
electrons, and is therefore quite model independent. Existing beta
decay experiments extract limits on an effective electron neutrino
mass $m_{\nu_e}^2 = \sum_i |U_{ei}|^2 m_i^2$ \cite{Farzan:2002zq},
provided that the neutrino masses are smaller than the detector
energy resolution. Currently the most stringent bounds on
$m_{\nu_e}^2$ are $(2.3)^2~\rm{eV}^2$ at $95\%$ confidence from the
Mainz experiment \cite{mainz} and $(2.5)^2~\rm{eV}^2$ at $95\%$
confidence from the Troitsk experiment \cite{troitsk}.  In the next
few years the Katrin experiment should exceed these limits by nearly
two orders of magnitude, probing down to $(0.2)^2~\rm{eV}^2$ at the
$90\%$ confidence level \cite{katrin}.  One might naively compute
this effective mass for the ``best fit'' mixing parameters obtained
in the previous section. In this case, we expect a keV seesaw
neutrino to contribute to $m^2_{\nu_e}$ by a huge amount $\delta
m_{\nu_e}^2 = U_{e6}^2 m_6^2 \sim \frac{m_l}{m_6} m_6^2 = m_l m_6$
\cite{SeeSaw_LSND} so that it would be excluded by the current
precision measures of tritium beta-decay for $m_l\gtrsim
10^{-3}$~eV. This is clearly an incorrect treatment of the physics.
As pointed out in, for example,  \cite{Farzan:2001cj}, the existence
of a heavy neutrino state would produce a kink in the electron
energy spectrum of size $|U_{ei}|^2$ at an energy $E_0 - m_i$ as
well as a suppression of events at the endpoint of order $1 -
|U_{ei}|^2$. Here $U_{ei}$ ($i=4,5,6$) is the mixing between the
electron neutrino flavor eigenstate and the heavy mass eigenstate,
while $E_0 = 18.6~\rm{keV}$ is the endpoint energy of tritium
beta-decay.

\begin{figure}
\begin{center}
\includegraphics[scale=0.70]{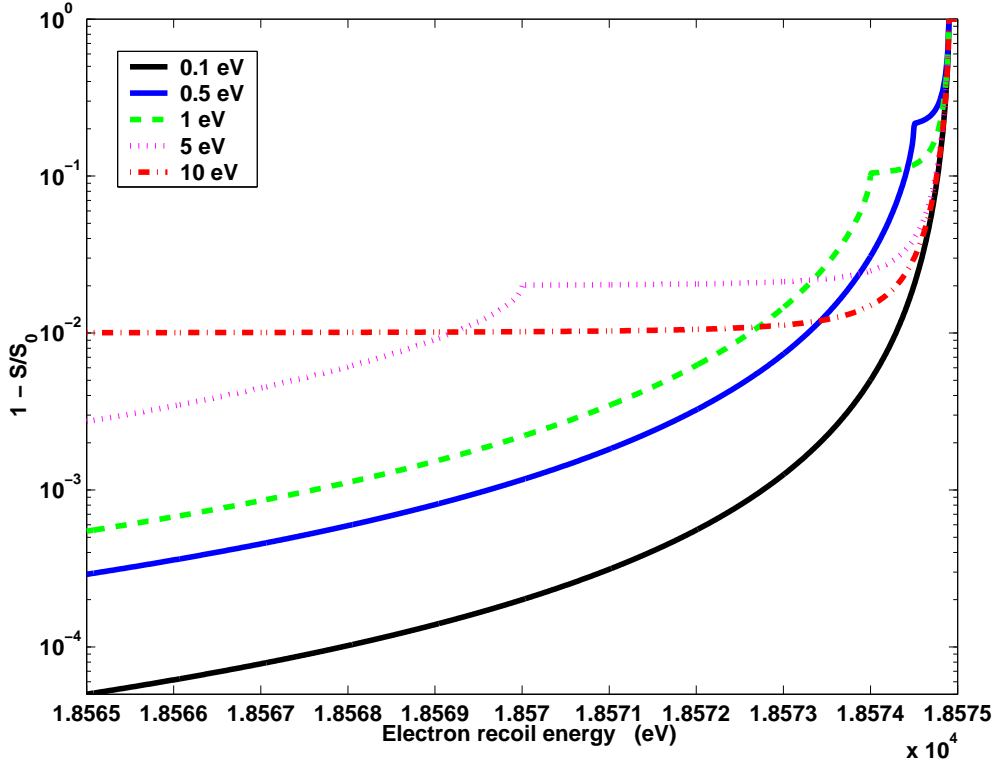}
\caption{$1-S/S_0$ as a function of the $\beta$-ray energy, where
$S$ is the $\beta$-ray energy spectrum obtained assuming three
mostly active, degenerate neutrinos with mass $m=0.1$~eV and one
mostly sterile neutrino $\nu_i$ with $m_i=0.1,0.5,1,5,$ and $10$~eV.
The mixing angle is given by  $U_{ei}^2=m/m_i$. $S_0$ is the
spectrum associated with massless neutrinos. See text for details.}
\label{fig:spectra}
\end{center}
\end{figure}

Fig.~\ref{fig:spectra} depicts $1-S/S_0$, where $S$ is the
$\beta$-ray energy spectrum obtained assuming three mostly active,
degenerate neutrinos with mass $m=0.1$~eV and one mostly sterile
neutrino $\nu_i$ with various masses $m_i$  and mixing  angle
$U_{ei}^2=m/m_i$, while $S_0$ is the spectrum associated with
massless neutrinos. One can readily observe ``kinks'' in the
spectrum above $m_i$. For $\beta$-energies above $E_0-m_i$, the
impact of the sterile state is to ``remove'' around $1 - |U_{ei}|^2$
of the $\beta$-rays from spectrum. This is most significant between
$E_0-m_i$ and $E_0$ minus the mass of the active neutrinos. For
energies below $E_0-m_i$, the spectrum agrees with that obtained
from the emission of one effective neutrino with mass-squared
$m_{\nu_e}^2$.

We estimate the sensitivity of future tritium beta-decay experiments
to the emission of one heavy state by considering the ratio between
the number of electrons with energies above $E_0-\Delta E$ in the
case of one heavy massive neutrino $\nu_i$ and in the case of
massless neutrinos,
\begin{equation}
 R(U_{ei},M_R) = \left[|U_{ei}|^2 \int_{E_0 - \Delta E}^{E_0}
 dE \frac{dN}{dE}(m_i) + \left(1 - |U_{ei}|^2 \right)\int_{E_0 - \Delta E}^{E_0} dE
 \frac{dN}{dE}(0)\right]\left/\int_{E_0 - \Delta E}^{E_0} dE
 \frac{dN}{dE}(0)\right.,
 \label{equ:R}
 \end{equation}
 where $dN/dE$ is the energy distribution of $\beta$-rays, which depends on the neutrino mass $m_i$.
This expression can be easily generalized for more than one heavy neutrino.  The advantage of using the ratio above is that potential systematic uncertainties and
normalization effects can be safely ignored.
An experiment is sensitive to a massive neutrino state if it can
distinguish $R$ from unity, a determination that should be limited
by statistics due to the very low $\beta$-ray flux in the high-energy tail of the
electron spectrum.

\begin{figure}
\begin{center}
\includegraphics[scale=0.70]{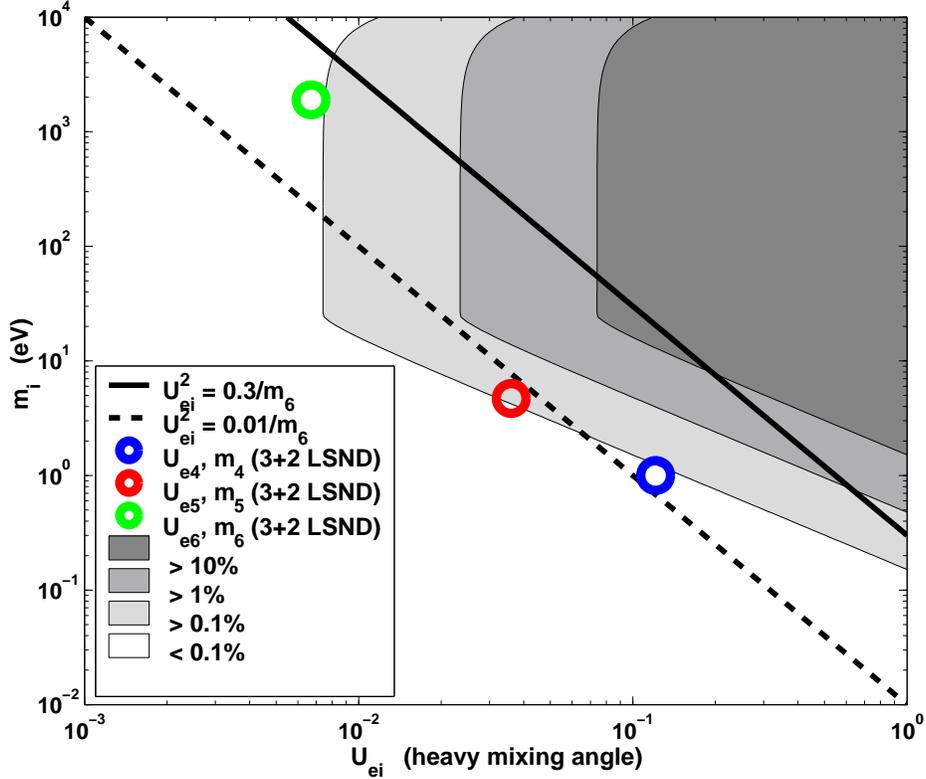}
\caption{Contour plot of constant $R$, as defined by
Eq.~(\ref{equ:R}), assuming an energy window $\Delta E =
25~\rm{eV}$.  The solid (dashed) line corresponds to
$\sqrt{m_l/m_i}$, a naive upper bound for $|U_{ei}|$, for
$m_l=0.3$~eV (0.01~eV). The circles correspond to $U_{ei}$ for the
three mostly sterile states obtained by our ``fit'' to other
neutrino data, Eq.~(\ref{U_32}). See text for details.}
\label{fig:Katrin}
\end{center}
\end{figure}

In order to compute $R$, we use an analytic expression for
Eq.~(\ref{equ:R}), which exists provided that one neglects nucleon
recoil in the decay. Fig.~\ref{fig:Katrin} depicts constant
$R$-contours in the $|U_{ei}|\times m_i$ plane. Contours were
computed for $\Delta E=25~\rm{eV}$, in order to allow one to easily
compare our results with the sensitivity estimates of the Katrin
experiment. After data-taking, Katrin is expected to measure $R$ at
the 0.1\%--1\% level (lightest grey region). Its sensitivity is
expected to be $\sqrt{m^2_{\nu_e}}>0.2$~eV at the 90\% confidence
level. This can be extracted from the plot by concentrating on the
$U_{ei}=1$ line.  Note that while the expected energy resolution for
Katrin is of order 1~eV, the expected number of signal events above
$E_0-1$~eV is minuscule (both in absolute terms and compared with
expected number of background events), so that most of the
sensitivity to nonzero neutrino masses comes from analyzing the
shape of the electron spectrum in the last tens of electron-volts. A
larger ``window'' would suffer from increased systematic
uncertainties, so that $\Delta E\sim 25$~eV is representative of
Katrin's optimal reach \cite{katrin}.

The shape of the constant $R$ contours is easy to understand. As
already discussed, for $m_i>\Delta E$, the effect of the
right-handed neutrinos is to reduce the spectrum in an energy
independent way by $1-|U_{ei}|^2$, while for $m_i<\Delta E$, states
with the same effective mass-squared $m_i^2|U_{ei}|^2$ produce the
same effect in tritium beta-decay so that the diagonal lines
coincide with lines of constant $m_i^2|U_{ei}|^2$.

A more sensitive approach would be to ``bin'' the last tens of eV of
the ``data'' into 1~eV bins, and fit the distribution to a massless
neutrino hypothesis. For the values of the parameters in which we
are interested, we find that one 25~eV bin yields roughly the same
sensitivity to nonzero neutrino masses as twenty five 1~eV bins for
large masses and small mixing angles. For smaller masses and larger
mixing angles, a ``binned'' analysis should be sensitive to effects
which are localized in individual bins (such as ``kinks'').  Another
recent estimate of the sensitivity of Katrin to heavy, sterile
neutrinos can be found in  \cite{smirnov_renata}.

Fig.~\ref{fig:Katrin} also depicts the loose upper bound for
$U_{ei}=\sqrt{m_l/m_i}$ as a function $m_i$, for $m_l=0.32$~eV and
$m_l=0.01$~eV. For $m_l\gtrsim 0.1$~eV, Katrin should be sensitive
to $M_R\lesssim1$~keV, while for $m_l\gtrsim 0.01$~eV (the solar
mass scale) Katrin should be sensitive to $M_R\gtrsim10$~eV and
$M_R\lesssim 100$~eV, where here we assume that all right-handed
neutrino masses are of order $M_R$. In the case of seesaw parameters
that fit the LSND data with a 3+2 neutrino spectrum (see
Eq.~(\ref{U_32})), expectations are high as far as observing a
kinematical neutrino mass effect at Katrin, in spite of the fact
that the fit to LSND data requires $U_{e4}$ and $U_{e5}$ to be
``abnormally'' low. Fig.~\ref{fig:Katrin} depicts $U_{ei}$ and $m_i$
values for the heavy neutrinos (open circles). The contribution of
the heaviest of the two LSND-related sterile neutrinos is of order
the Katrin sensitivity, while the active contribution itself, which
leads to $m^2_{\nu_e}=\sum_{i=1,2,3}|U_{ei}^2m_i^2|\sim m_l^2$ is
already within the Katrin sensitivity range, given that large
$m_l>0.22$~eV values are required by our 3+2 LSND ``fit''. The
effect of $\nu_6$ is small if $m_6$ is larger than 1~keV (required
if one takes the ``pulsar kicks'' hint into account), but would be
very significant if $m_6$ were less than 1~keV.

\subsection{Neutrinoless Double-Beta Decay}

If the neutrinos are Majorana fermions -- as predicted in the case
of interest here -- lepton number is no longer a conserved quantity.
The best experimental probe of lepton number violation is the rate
for neutrinoless double-beta decay. This process, which violates
lepton number by two units, is currently the subject of intense
search \cite{bb0n_future}. If neutrino masses are the only source of
lepton number violation, the decay width for neutrinoless
double-beta decay is
\begin{equation}
\Gamma_{0\nu\beta\beta}\propto \left|\sum_i U_{ei}^2 \frac{m_i}{Q^2+m_i^2}{\cal M}(m_i^2,Q^2)\right|^2,
\label{eq:mbb}
\end{equation}
where $\cal M$ is the relevant nuclear matrix element and $Q^2\sim
50^2$~MeV$^2$ is the relevant momentum transfer. In the limit of
very small neutrino masses ($m_i^2\ll Q^2$),
$\Gamma_{0\nu\beta\beta}$ is proportional to an effective neutrino
mass $|m_{ee}|$,
\begin{equation}
m_{ee} = \sum_i^n U_{ei}^2 m_i . \label{equ:m_ee}
\end{equation}
The sum is over all light neutrino mass eigenstates. In the case of
a low-energy seesaw, when all $m_i$, $i=1,\ldots,6$ are much smaller
than $Q^2$, it is easy to see that $m_{ee}$ vanishes
\cite{SeeSaw_LSND}. The reason for this is that, in the weak basis
we are working on (diagonal charged-lepton and charged
weak-current), $m_{ee}$ is the $ee$-element of the neutrino mass
matrix, as defined in Eq.~(\ref{eq:lagrangian}). One can trivially
check that, by assumption, not only does $m_{ee}$ vanish, but so do
all other $m_{\alpha\beta}$, $\alpha,\beta=e,\mu,\tau$. Note that
this result does not depend on the fact that we have been assuming
all elements of the neutrino mass matrix to be real
\cite{Kayser_CP}.

For heavy $\nu_i$ neutrinos, $U_{ei}^2m_i$ no longer captures the
dependency of $\Gamma_{0\nu\beta\beta}$ on the exchange of $\nu_i$.
For $m_i^2\gg Q^2$, instead, the dependency on neutrino exchange is
proportional to $U_{ei}^2 /m_i$. If this is the case, the overall
contribution (including all heavy and light states) is no longer
proportional to $m_{ee}$ but, instead, can be qualitatively
expressed as a function of an effective $m^{\rm eff}_{ee}$,
\begin{equation}
m_{ee}^{\rm eff}\equiv Q^2\sum_i \frac{U_{ei}^2 m_i}{Q^2+m_i^2}.
\end{equation}
The approximation $\Gamma_{0\nu\beta\beta}\propto |m_{ee}^{\rm
eff}|$ is good as long as one can neglect the dependency of ${\cal
M}$ on $m_i$ and is not expected to be a great approximation when
$m_i^2\sim Q^2$. Nonetheless, $m_{ee}^{\rm eff}$ still qualitatively
captures the behavior of $\Gamma_{0\nu\beta\beta}$ as a function of
the sterile neutrino masses and studying its behavior is sufficient
for our ambitions in this discussion.

\begin{figure}
\begin{center}
\includegraphics[angle = 270,scale=0.6]{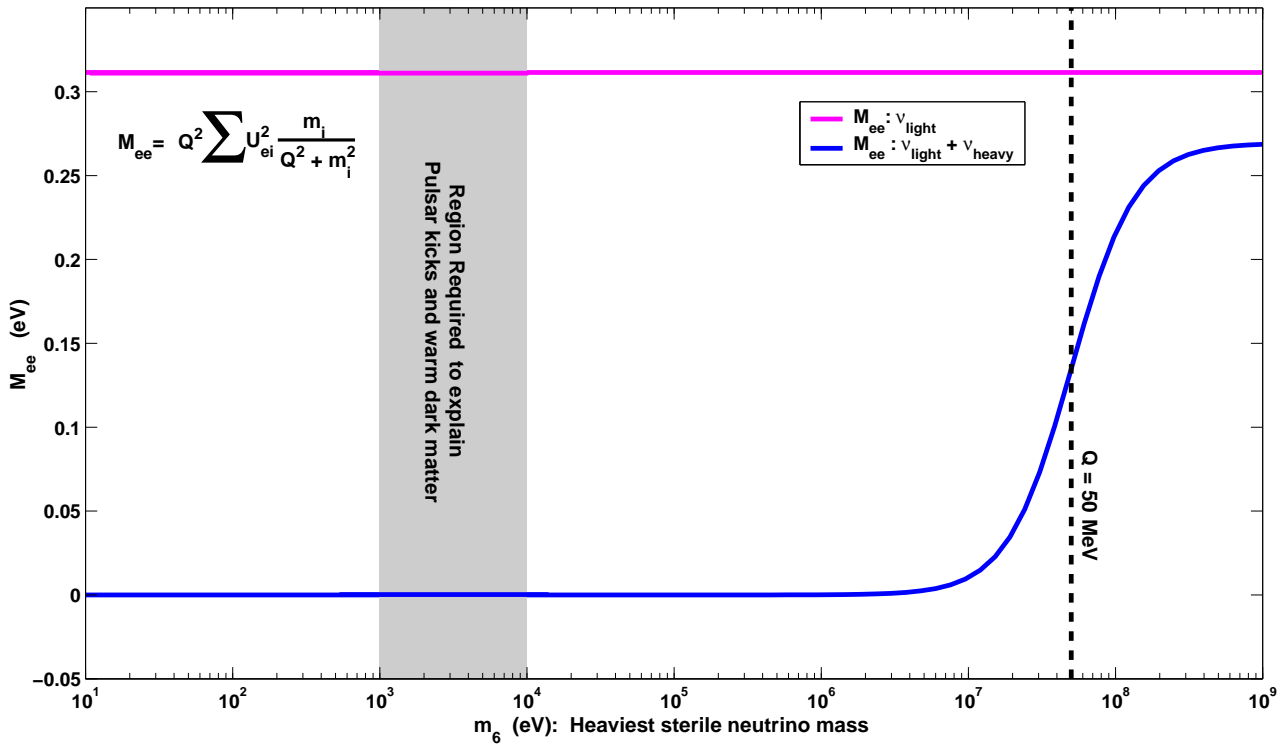}
\caption{Effective $m_{ee}$ for
neutrinoless double-beta decay as a function of $m_6$, the heaviest right-handed neutrino mass,   assuming the existence of only light, active, neutrinos (magenta
curve), with a degenerate mass spectra, and for our ``best fit''
$3+2$ LSND sterile neutrino solution (blue curve).  See text for details. Also indicated is
the parameter region preferred by astrophysical hints of sterile neutrinos. We assume $Q =  50~\rm{MeV}$.
In the case of a low-energy seesaw, $m_{ee}$ vanishes as long as $m_6\ll Q$.}
\label{fig:0nubb}
\end{center}
\end{figure}

Fig.~\ref{fig:0nubb} depicts $m_{ee}^{\rm eff}$ for our ``best fit''
$3+2$ LSND solution (see Sec.\ref{subsec:LSND}), as a function of
the unconstrained $m_6$. As advertised, $m_{ee}^{\rm eff}$ vanishes
for $m_6^2\ll Q^2$. The figure also depicts the ``active only''
value of $m_{ee}^{\rm active}=\sum_{i=1,2,3}U_{ei}^2m_i$. Even in
the limit $m_6^2\gg Q^2$, there is still partial cancelation between
the mostly active and mostly sterile ``LSND'' states.   This is a
feature of the Lagrangian we are exploring here, and is not in
general observed in other scenarios with light sterile neutrinos
tailor-made to solve the LSND anomaly.

Currently, the most stringent limits on this effective mass comes
from the Heidelberg-Moscow experiment \cite{bb0n_present} where they
find $m_{ee} < 0.91~\rm{eV}$ at $99\%$ confidence. In the near
future, experiments aim to reach down to $m_{ee}$ values close to
$10^{-2}$~eV \cite{bb0n_future}. A signal would rule out a seesaw
scale below tens of MeV. On the other hand, if we were to conclude
that the neutrino masses are quasi-degenerate (through, say, a
signal in tritium beta-decay) and if the LSND 3+2 solution were
experimentally confirmed, a vanishing result for $m_{ee}$ could be
considered strong evidence for a very small seesaw scale. On the
other hand, if this were the case ($m_{ee}$ zero, large active
neutrino masses), it would also be very reasonable to conclude that
neutrinos are Dirac fermions. Distinguishing between the two
possibilities would prove very challenging indeed. It is curious
(but unfortunate) that in a low energy seesaw model, the neutrinos
are Majorana fermions, but all ``standard'' lepton-number violating
observables vanish, given that their rates are all effectively
proportional to $m_{\alpha\beta}$!

\section{Concluding Remarks}
\label{sec:Conclusion}

The ``New Standard Model'' (equal to the ``old'' Standard Model plus
the addition of three gauge singlet Weyl fermions) is, arguably, the
simplest extension of the SM capable of accommodating neutrino
masses. This Lagrangian contains a new dimension-full parameter: the
right-handed neutrino mass scale $M$, which must be determined
experimentally. Unlike the Higgs mass-squared parameter, all $M$
values are technically natural given that the global symmetry of the
Lagrangian is enhanced in the limit $M\to 0$.

Very large $M$ values are \emph{theoretically} very intriguing, and
have received most of the attention of the particle physics
community. There are several strong hints that new phenomena are
expected at the electroweak breaking scale $\sim 10^3$~GeV, the
grand unified scale $\sim 10^{15-16}$~GeV, and the Planck scale
$\sim 10^{18-19}$~GeV, and it is tempting to associate $M$ to one of
these energy scales. Furthermore, large $M$ values provide an
elegant mechanism for generating the matter antimatter asymmetry of
the Universe \cite{leptogenesis}. Of course, large $M$ values are
experimentally very frustrating. It may ultimately prove impossible
to experimentally verify whether the New Standard Model is really
the correct way to describe Nature.

Here, we explore the opposite end of the $M$ spectrum, $M\lesssim
1$~keV. Such values are \emph{phenomenologically} very intriguing,
given that small $M$ values imply the existence of light sterile
neutrinos that mix significantly with the active neutrinos and can
potentially be directly observed. Furthermore, there are several
experimental and astrophysical phenomena that are best understood if
one postulates the existence of light, moderately mixed sterile
neutrino. We find that by requiring all three right-handed neutrino
masses to be less than a few keV we can simultaneously explain all
neutrino oscillation data, including those from LSND, explain the
large peculiar velocities of pulsars, and accommodate the production
of heavy elements in supernova environments. Our fit also provides
constraints for the ``active'' neutrino oscillation parameters, most
strongly to the lightest active neutrino mass. All successful
parameter choices that accommodate the LSND data require $m_l$ to be
``large'' ($m_l\gtrsim 0.1$~eV), and the ``best fit'' requires all
active neutrino masses to be quasi-degenerate.

On the negative side, low $M$ values are, theoretically, rather
puzzling. In order to obtain the observed light neutrino masses,
neutrino Yukawa couplings are required to be much smaller than the
electron Yukawa coupling, and it is tempting to believe that such
small numbers are proof that a more satisfying understanding of
fermion masses must exist. Furthermore, thermal leptogenesis is no
longer an option (see, however, \cite{Shaposhnikov:2006xi}).
Finally, the fact that $M$ is naively unrelated to other mass scales
can also be perceived as disheartening, but in our opinion, should
be interpreted as evidence that there is more to the lepton sector
than meets the eye.

Regardless of one's preference for a high or low seesaw energy
scale, and independent of whether the data from LSND and the
astronomical observables discussed above have anything to do with
sterile neutrinos, our main point is that the determination of  $M$
is an \emph{experimental} issue. In the near/intermediate future,
low energy seesaw scales will be probed by several experiments; most
importantly measurements of the end-point of tritium beta-decay, the
MiniBooNE experiment, searches for neutrinoless double-beta decay
and, if we get lucky, the detection of neutrinos from a nearby
supernova explosion. We find, for example, that Katrin should be
sensitive to seesaw energy scales below tens of keV if all
right-handed neutrino masses are similar, while null results from
MiniBooNE would severely constrain right-handed neutrino masses
below several eV. We conclude by pointing out that larger (but still
``small'') values of $M$ are much harder to constrain. For GeV
sterile neutrinos, typical active--sterile mixing angles are
$U_{\alpha i}^2\lesssim 10^{-10}$, probably too small to observe in
particle physics processes.  It is frustrating (and, we hope,
ultimately false) that we seem to be unable to experimentally
distinguish $M\sim 1$~GeV from $M\sim 10^{14}$~GeV\ldots

\section*{Acknowledgments}
We thank Irina Mocioiu for comments on how the pulsar kick requirements extend toward the $\Omega_s>0.3$ region, and Michel Sorel for discussions regarding 3+2 fits to the neutrino oscillation data. JJ would also like to thank the students and lecturers of the TASI 2006 summer school for useful discussions of this manuscript
and many related physical concepts. This work is sponsored in part by the DOE grant \# DE-FG02-91ER40684.

 \end{document}